# Kaggle Competition: Expedia Hotel Recommendations

**Gourav G. Shenoy, Mangirish A. Wagle, Anwar Shaikh**
Indiana University
Bloomington, IN, USA
`{goshenoy, mawagle, anshaikh}@indiana.edu`

## 1. Introduction

With hundreds, even thousands, of hotels to choose from at every destination, it's difficult to know which will suit your personal preferences. Expedia wants to take the proverbial rabbit hole out of hotel search by providing personalized hotel recommendations to their users. This is no small task for a site with hundreds of millions of visitors every month! Currently, Expedia uses search parameters to adjust their hotel recommendations, but there aren't enough customer specific data to personalize them for each user. In this project, we have taken up the challenge to contextualize customer data and predict the likelihood a user will stay at 100 different hotel groups.

The train/test datasets used for this project have been provided by Expedia, via Kaggle, and they contain 23 features capturing the logs of customer behavior. There is an additional dataset containing 149 features, which pertain to the hotel reviews made by users. The goal is to build a machine learning model to predict the booking outcome (hotel cluster) for a user event, based on their search and other attributes associated with that user event. As part of this project, we have implemented the following algorithms:

1. Naive Bayes
2. Decision Trees
3. K Nearest Neighbors
4. K Means Clustering
5. Multinomial Logistic Regression
6. Ensemble Learning Methods with k-NN and Decision Trees.

We have also compared our results with off-the-shelf algorithms provided by machine learning Python package, SciKit-Learn. Section (2) below talks more about the data-set provided by Expedia; Section (3) talks about the preliminary analysis we conducted; Section (4) gives more details about the algorithms implemented, along with their results; Section (5) compares our results with SciKit-Learn's off-the-shelf algorithms; And the details about our source code on Github is provided in Section (6);

## 2. The Expedia Dataset

Expedia provided data-set that captured the logs of user behavior. These include details about what the customers searched for, how they interacted with the search results - i.e.

whether they actually booked the hotel, or simply clicked to view details, whether or not the search result was a travel package, and so on. The goal is to predict which "hotel cluster" the user is likely to book, given his search details. These "clusters" have been created by Expedia based on some undisclosed in-house algorithms. But the intuition is that hotels belonging to a cluster are similar for a particular search - based on historical price, customer star ratings, geographical locations relative to city center, etc. These hotel clusters serve as good identifiers to which types of hotels people are going to book, while avoiding outliers such as new hotels that don't have historical data.

The training and testing data-sets are split based on time: training data from 2013 and 2014, while test data are from 2015. Training data includes all the users in the logs, including both click events and booking events. Test data only includes booking events. The table (1) below provides the schema of the train/test data-sets.

| Feature Name | Feature Description | Feature Data-Type |
|---|---|---|
| date_time | Time stamp | string |
| site_name | ID of the Expedia point of sale | int |
| posa_continent | ID of continent associated with site_name | int |
| user_location_country | The ID of country customer is located | int |
| user_location_region | The ID of region customer is located | int |
| user_location_city | The ID of city the customer is located | int |
| orig_destination_distance | Physical distance between a hotel and a customer at the time of search. | double |
| user_id | ID of user | int |
| is_mobile | 1 when a user connected from a mobile device, 0 otherwise | tinyint |
| is_package | 1 if the click/booking was generated as a part of a package, 0 otherwise | int |
| channel | ID of a marketing channel | int |
| srch_ci | Check-in date | string |
| srch_co | Check-out date | string |
| srch_adults_cnt | The number of adults specified in the hotel room | int |
| srch_children_cnt | The number of (extra occupancy) children specified in the hotel room | int |
| srch_rm_cnt | The number of hotel rooms specified | int |

| | in the search | |
|---|---|---|
| srch_destination_id | ID of the destination where the hotel search was performed | int |
| srch_destination_type_id | Type of destination | int |
| hotel_continent | Hotel continent | int |
| hotel_country | Hotel country | int |
| hotel_market | Hotel market | int |
| is_booking | 1 if a booking, 0 if a click | tinyint |
| cnt | Number of similar events in the context of the same user session | bigint |
| hotel_cluster | ID of a hotel cluster | int |

**Table-1:** Data fields in the train/test data-set

Expedia also provided with another data-set named "destinations.csv" which contains information related to hotel reviews made by users, and extracted them as features. This data has 149 features, and a reference-key-feature "srch_destination_id". The table (2) below provides the schema of the destinations data-set.

| Feature Name | Feature Description | Feature Data-Type |
|---|---|---|
| srch_destination_id | ID of the destination where the hotel search was performed | int |
| d1-d149 | latent description of search regions | double |

**Table-2:** Data fields in the destinations data-set

The train/test data-sets are pretty huge in terms of size. The train data-set contains nearly 300 million records, while the test data-set contains nearly 75 million records. The additional "destinations" data-set contains nearly 62k records.

## 3. Preliminary Analysis

In order check whether there is redundancy in the features we created correlation matrix and plot for the data which is depicted as follows in Figure-1:

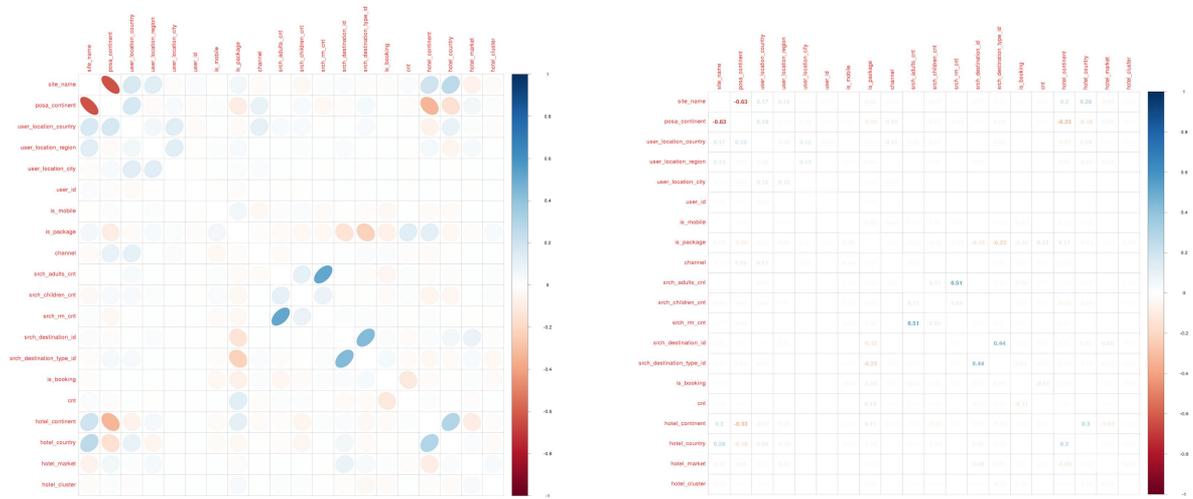

**Figure-1:** Correlation Matrix

Looking at the plots and numbers we conclude that there are no significant correlation between features except some demographic features like user country and site (.uk, .ca etc) he is accessing which is ought happen.

As we have mentioned the data is having 300 million examples. It has user's booking as well as click data. In order to make data easy to handle we filtered to get only booking data. That reduced data to 3 million examples still large number of examples.

In data there are 100 clusters which are precalculated by expedia. We first took a look at the class distribution by histogram which is as follows

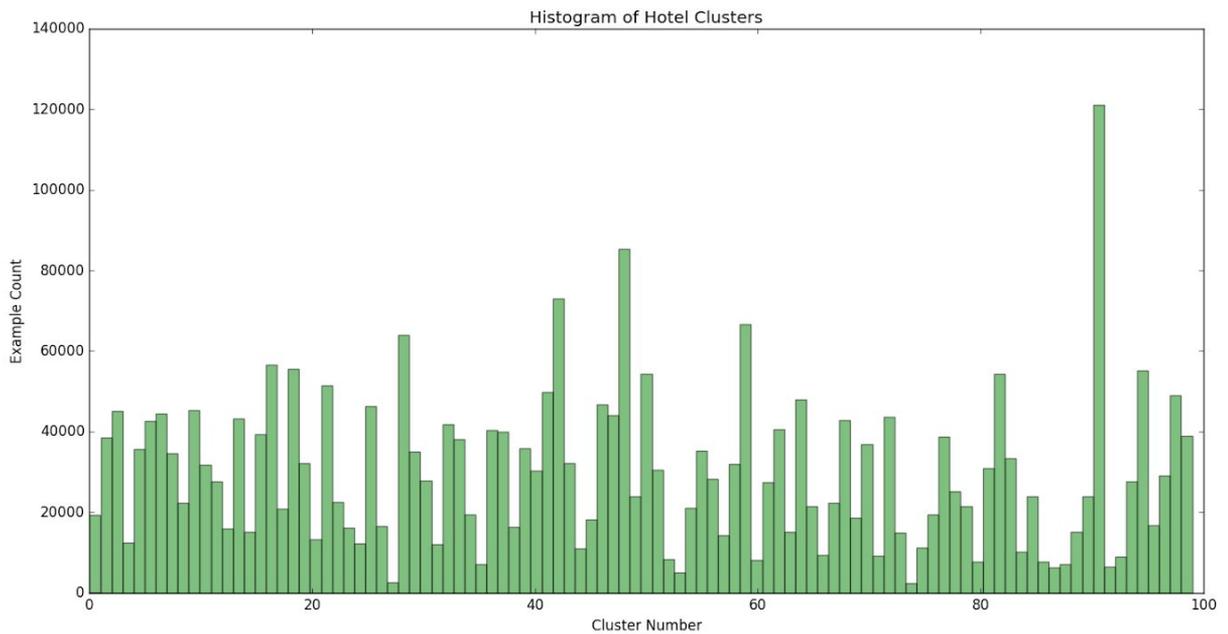

**Figure-2:** Histogram of hotel clusters

As we see the data is well distributed over the all 100 classes and there is skewness in the data.

We foresee a potential problem with so many classes in the data having 100 classes for the classification problem would result in decrease in performance of the model. Hence we have used a k-means clustering to group the data together. We created clusters with k=5, 10, 20 and 50. We will show the their impact on the result in our analysis.

For this purpose we used the k-means algorithm in Scikit-Learn library with k-means++ initialization. Following are graphs which represents how the created clusters are related we present stack bar graph for 5 clusters vs original hotel clusters 100 and 5 clusters vs created 10 clusters.

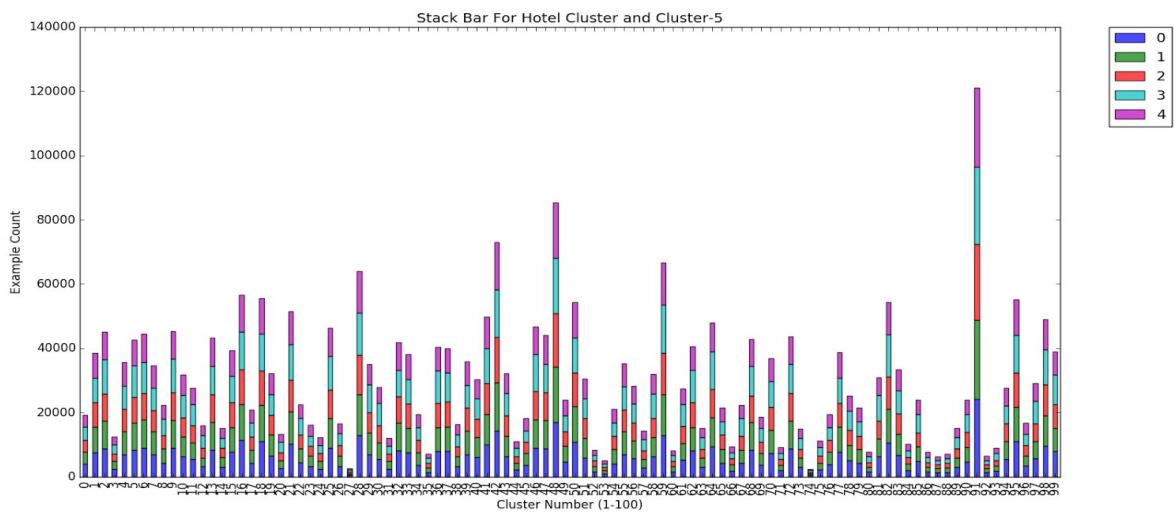

**Figure-3:** Stack bar for hotel cluster & cluster-5

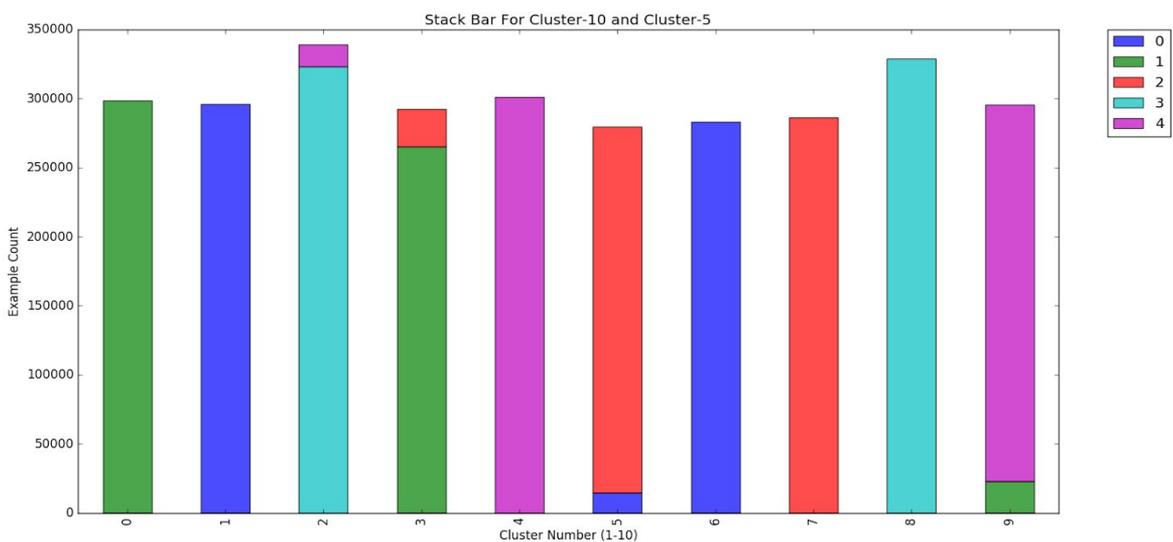

**Figure-4:** Stack bar for cluster-10 and cluster-5

## 4. Algorithms Implemented

Following are the data pre-processing steps that were executed before providing the data to the actual algorithm implementation:

1. Merging the datasets using the srch_destination_id with the destinations dataset.
2. Discretize the continuous data. E.g. changing the datetime to discrete values as follows, 10-04-2016 was split into two features, Month: 10 and Year: 2016.

We implemented the following set of algorithms to analyze the expedia data. Also, for training the models using below algorithms, we used the "clustered" data-set created as part of the preliminary analysis step. **Note:** we created 4 clusters of size(s) 5, 10, 20, 50, and 100 out of the original data-set with 100 class labels. We planned on comparing the accuracies of the models with these cluster sizes.

### 4.1. Naive Bayes

#### 4.1.1 Configuration & Results

We trained the naive-bayes classifier with the following 5 configurations (related to data-set cluster sizes), and the corresponding accuracies:

| Cluster Size | Accuracy |
|---|---|
| 5 | 33.01% |
| 10 | 18.32% |
| 20 | 6.88% |
| 50 | 2.09% |
| 100 | 1.85% |

**Table-3:** Cluster Size and Accuracy

#### 4.1.2 Graph Plot

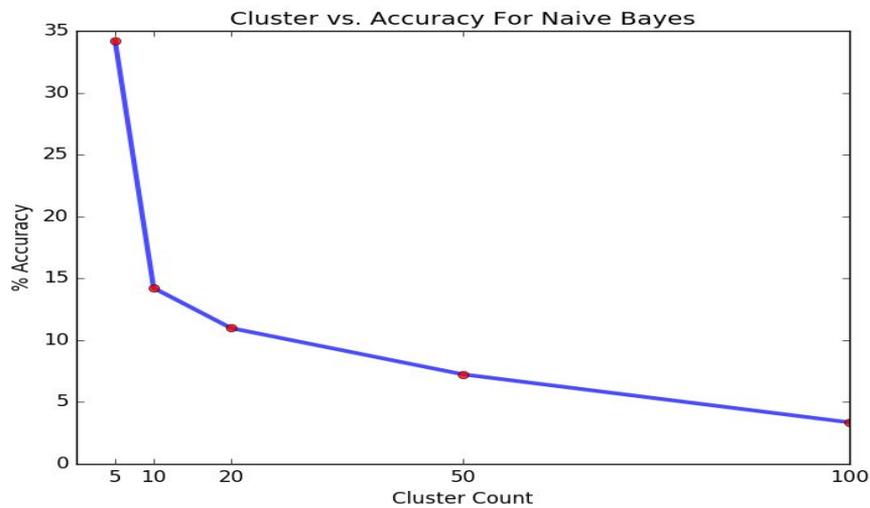

**Figure-5:** Cluster vs Accuracy for Naive Bayes

**4.1.3 Analysis**

We observed that the accuracy of Naive-Bayes is highest for cluster size 5, and 10. But as we increase the cluster size, the accuracy reduces. There is a steep decline in the accuracy, which might mean that given lesser options to choose from (eg: 5 classes), the chances of misclassification is smaller compared to say 100 classes.

**4.2. Multinomial Logistic Regression**

**4.2.1 Configuration & Results**

We have implemented multinomial logistic regression as data is numeric and have many classes. We used the clusters that we have created and ran the algorithm with k-fold cross validation with k = 5, 10.

| K-Fold Validation | Accuracy |
|---|---|
| *Cluster Size = 5* | |
| 5 | 34.23% |
| 10 | 37.13% |
| *Cluster Size = 10* | |
| 5 | 14.22% |
| 10 | 15.83% |
| *Cluster Size = 20* | |
| 5 | 10.98% |
| 10 | 11.82% |
| *Cluster Size = 50* | |
| 5 | 7.24% |
| 10 | 7.93% |
| *Cluster Size = 100* | |
| 5 | 3.36% |
| 10 | 3.41% |

**Table-4:** Multinomial Logistic Regression Summary

**4.2.2 Graph Plots**

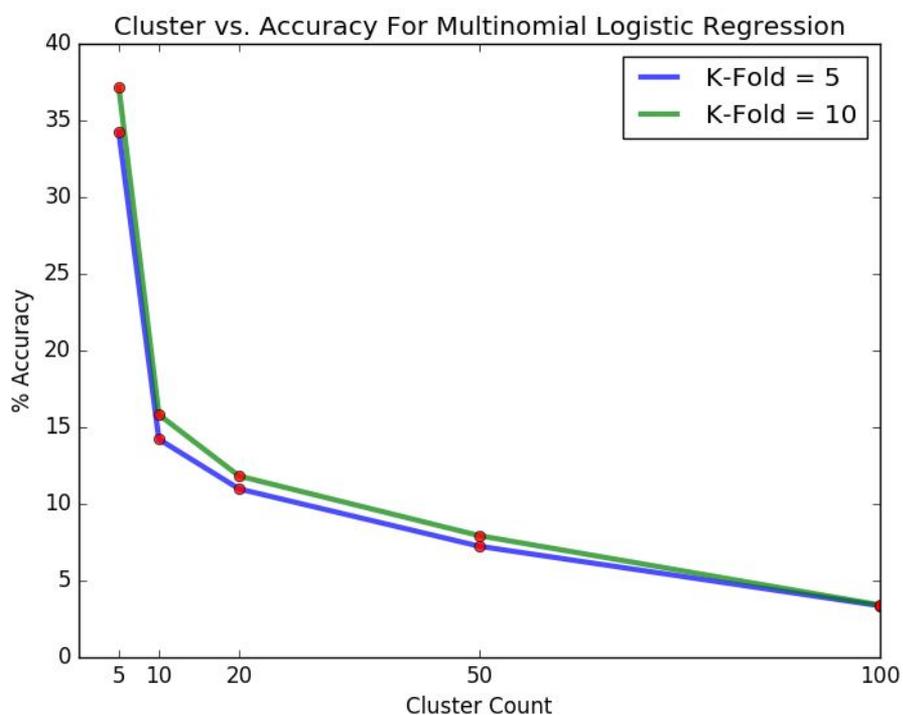

**Figure 6:** Cluster vs Accuracy for Multinomial Logistic Regression

**4.2.3 Analysis**

We again observe here that the accuracy is highest for cluster sizes 5 and 10, whereas for greater cluster sizes, the accuracy seems to reduce. Based on this observation, we can run our subsequent algorithms for size 5, 10 only.

**4.2. Decision Trees**

**4.3.1 Configuration & Results**

We noticed that only cluster 5 and 10 are giving good accuracies and others are just following the pattern. Also, since it takes a long time to train the model, we decided to only execute the algorithm for clusters 5 and 10 respectively.

| Tree Depth | Accuracy |
|---|---|
| *Cluster Size = 5* | |
| 3 | 18.77% |
| 5 | 19.23% |
| 10 | 19.11% |
| *Cluster Size = 10* | |
| 3 | 6.08% |
| 5 | 7.76% |
| 10 | 10.03% |

**Table-5:** Decision Trees Summary

**4.3.2 Graph Plot**

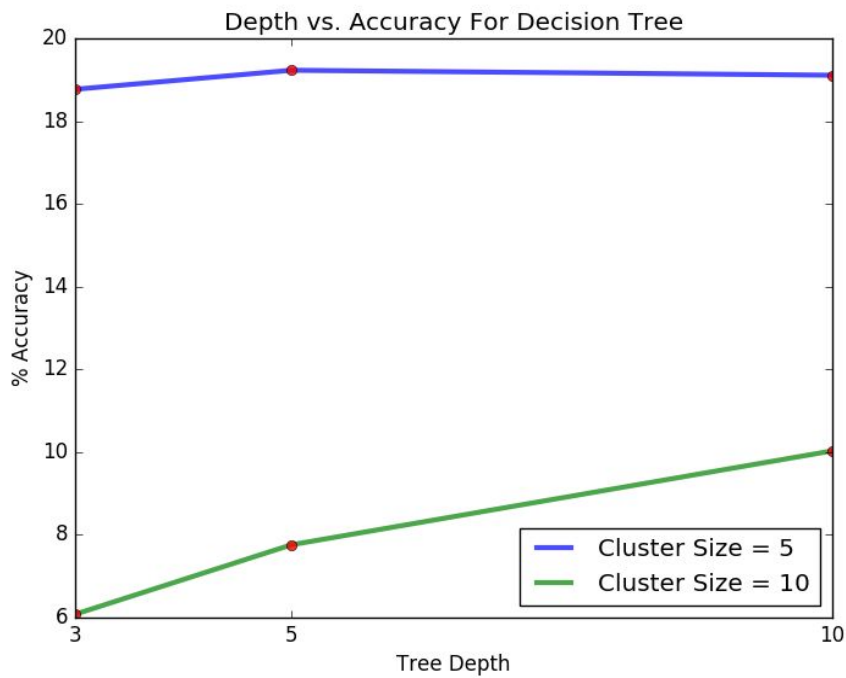

**Figure-7 :** Tree depth vs Accuracy For Decision Tree

**4.3.3 Analysis**

We notice that the decision tree algorithm performs average for this data-set. One reason could be that the existence of continuous data.

### 4.4. K-Nearest Neighbors

#### 4.4.1 Configuration & Results

We trained the k-NN classifier model for 2 values of k, i.e. 10, and 20 respectively. A point to note here is that the data-set used for training & testing is sampled data; as the original data-set was too large and the program took extremely long to finish. Hence we decided to randomly sample data points from the original data-set and learn the model.

| K-value | Accuracy |
|---|---|
| 10 | 5.61% |
| 20 | 8.09% |

**Table-6:** K-NN Summary

#### 4.4.3 Analysis

In our analysis we find that k-nearest neighbour is relatively weak classifier for the given problem. We tried with different k-values but accuracy does not get any better. We did not use our manually created clusters as they are created with very same process that k-nn is utilizing for the classification.

### 4.5. Ensemble Methods

#### 4.5.1 Configuration & Results

With the initial classification analysis of the data, the accuracy was found to be very low. Hence we chose to opt for learning model with ensemble learning methods to improve accuracy. We learnt the bagging and AdaBoost ensemble methods, using Decision Trees. **Note:** We have only considered data-sets with cluster size 5, 10 only.

| Tree Depth | Number of Bags | Accuracy |
|---|---|---|
| *Bagging, Cluster Size = 5* | | |
| 5 | 5 | 20.09% |
| 5 | 10 | 23.42% |
| 10 | 5 | 23.01% |
| 10 | 10 | 25.96% |
| *Bagging, Cluster Size = 10* | | |
| 5 | 5 | 11.53% |
| 5 | 10 | 14.11% |
| 10 | 5 | 14.43% |
| 10 | 10 | 17.61% |
| *AdaBoost, Cluster Size = 5* | | |
| 3 | 10 | 42.55% |
| 3 | 20 | 46.04% |
| 3 | 40 | 52.36% |
| *AdaBoost, Cluster Size = 10* | | |
| 3 | 10 | 31.61% |
| 3 | 20 | 37.23% |
| 3 | 40 | 37.95% |

**Table-7:** Ensemble Methods Summary

**4.5.3 Analysis**

In our analysis we came to know why boosting is known as powerful ensemble method. As we can see adaboost simply outperforms every other classifier. It results ~52% of accuracy for cluster size =5, which is very good looking at what and how data is. On other hand no other classifier is able cross 40% accuracy.

## 5. Comparison with off-the-shelf algorithms

We have used scikit-learn library for comparing and evaluating our algorithm implementations. We compared Logistic Regression, Decision Tree and adaboost algorithms. We find that these algorithms perform better than when running with larger size clusters they outperform our algorithms whereas in case of smaller size clusters like 5 and 10 results are comparable.

| Algorithm | Scikit-Learn | | Our Implementation | |
|---|---|---|---|---|
| | *Cluster = 5* | *Cluster = 10* | *Cluster = 5* | *Cluster = 10* |
| **Logistic Regression** | 40.11 | 38.51 | 37.12 | 34.23 |
| **Decision Tree** | 23.62 | 13.27 | 19.23 | 10.03 |
| **Adaboost** | 60.21 | 51.23 | 52.36 | 37.95 |

**Table-8:** Scikit-learn vs Our Implementation Comparison

## 6. Source Code (GitHub) & Technologies

We have maintained a GitHub repository for all implementations related to this project. All the changes have been pushed to the following repository:

https://github.com/gouravshenoy/ExpediaHotelRecommendation

We have used Python as our development language, and the interpreter is Python 3.5.2. The following additional package is used for helping with the data-processing, and graph plotting tasks:

- Pandas - for data processing.
- Matplotlib - for plotting the graphs

The preliminary analysis - including plotting the correlation matrix, filtering the data-set, etc - was performed using R programming language. Since the data-set was pretty huge, we needed to run the algorithms on machines other than our local laptops. We used the following compute resources to run the algorithms, perform analysis, and capture the results:

- Hulk - Linux based remote supercomputer provided by IU having 32-cores, and 512GB memory.
- Amazon EC2 - Centos07 instance with 8-cores, and 32GB memory.

## 7. Conclusion

The Expedia Hotel Booking dataset was analyzed by various machine learning algorithms that helped us come up with classification models for the Hotel Reservation System. The dataset has multiple classes without any significant perceived pattern that relates them to the features. This initially made it difficult to achieve reasonable accuracy. After we applied Clustering and Ensemble methods, we could achieve noticeable increase in accuracy. The highest observed was by the Multinomial Logistic Regression, as it handles the numeric data efficiently. Other algorithms used like Decision tree are known to work well in case of Discrete data.

The randomness induced by the real world events makes it challenging to learn patterns and this calls for extra curation of the data using various considerations.